\documentclass[twocolumn,english,prd,superscriptaddress,aps,showpacs,showkeys,amsmath,amssymb,floatfix,nofootinbib]{revtex4-1}
\usepackage[pdftex]{graphicx}
\usepackage[pdftex]{color}
\usepackage[colorlinks,bookmarks]{hyperref}
\definecolor{linkblue}{rgb}{0,0,0.8}
\definecolor{linkgreen}{rgb}{0,0.5,0}
\definecolor{darkgreen}{rgb}{0,0.4,0}
\definecolor{purple}{rgb}{0.7,0.0,0.4}
\hypersetup{linkcolor=linkblue, citecolor=purple, urlcolor=linkblue}

\usepackage{nicefrac}

\usepackage[T1]{fontenc}
\usepackage[utf8]{inputenc}
\usepackage{lmodern}
\usepackage{babel}
\usepackage{bm}
\usepackage{verbatim}
\usepackage[us]{datetime}
\usepackage{enumerate}

\makeatletter
\newcommand{\vast}{\bBigg@{4}}
\newcommand{\Vast}{\bBigg@{5}}
\makeatother

\def\rd{{\rm d}}

\newcommand{\omegam}{\Omega_{m0}}
\newcommand{\sig}{\sigma_8}
\newcommand{\omegal}{\Omega_{\Lambda0}}

\newcommand{\sint}{\sigma_{\rm int}}

\newcommand{\sinti}{\sigma_{{\rm int},i}}

\newcommand{\sintj}{\sigma_{{\rm int},j}}

\newcommand{\dd}{\textrm{d}}

\newcommand{\tgl}{\textsc{turboGL}\ }

\begin{document}

\title{Accurate weak lensing of standard candles. II. Measuring \boldmath $\sigma_8$ with Supernovae}

\author{Miguel Quartin}

\affiliation{Instituto de Física, Universidade Federal do Rio de Janeiro, CEP
21941-972, Rio de Janeiro, RJ, Brazil}

\author{Valerio Marra}

\affiliation{Institut für Theoretische Physik, Universität Heidelberg, Philosophenweg
16, 69120 Heidelberg, Germany}

\author{Luca Amendola}

\affiliation{Institut für Theoretische Physik, Universität Heidelberg, Philosophenweg
16, 69120 Heidelberg, Germany}

\begin{abstract}
    Soon the number of type Ia supernova (SN) measurements should exceed 100,000. Understanding the effect of weak lensing by matter structures on the supernova brightness will then be more important than ever. Although SN lensing is usually seen as a source of systematic noise, we will show that it can be in fact turned into signal. More precisely, the non-Gaussianity introduced by lensing in the SN Hubble diagram dispersion depends rather sensitively on the amplitude $\sig$ of the matter power spectrum. By exploiting this relation, we are able to predict constraints on $\sig$ of $7$\% ($3$\%) for a catalog of 100,000 (500,000) SNe of average magnitude error 0.12, without having to assume that such intrinsic dispersion and its redshift evolution are known \emph{a priori}. The intrinsic dispersion has been assumed to be Gaussian; possible intrinsic non-Gaussianities in the dataset (due to the SN themselves and/or to other transients) could be potentially dealt with by means of additional nuisance parameters describing higher moments of the intrinsic dispersion distribution function. This method is independent of and complementary to the standard methods based on CMB, cosmic shear or cluster abundance observables.
\end{abstract}

\keywords{Gravitational lenses, Observational cosmology, Supernovae, Large Scale Structure of the Universe}

\pacs{98.62.Sb, 98.80.Es, 97.60.Bw, 98.65.-r}

\maketitle

\section{Introduction}\label{intro}

Standard candles, in particular supernovae Ia (SNe), are one of the most important and reliable estimators
of distance in cosmology~\cite{Riess:1998cb,Perlmutter:1998np}. As is well known, the evidence for cosmological acceleration rests principally on
their properties and on their calibration. Since the discovery of acceleration, a large effort has been
devoted to testing and improving the calibration of the SNe and to correcting their light curves
in order to achieve data samples as free of systematics as possible~\cite{March:2011rv,Amendola:2012wc}.

Since their light comes from relatively high redshifts, SNe are expected to be lensed to some
extent by intervening matter along the line of sight. The correction induced by this effect is
normally subdominant but will become one of the major sources of uncertainty when richer and deeper
SN catalogs will be collected in the next years. The  Large Synaptic Survey Telescope (LSST) project plans for instance to collect up to half a million
SNe in ten years~\cite{Abell:2009aa}, a huge increase from the roughly 1000 SNe known so far.

The effect of gravitational lensing will in general change the intrinsic distribution function of the SN magnitudes, increasing the scatter and introducing some non-Gaussianity, if originally absent. In part I of our present investigation~\cite{Amendola:2013twa}, we have obtained the lensing variance, skewness and kurtosis of the SN distribution via sGL,  a fast simulation method developed in~\cite{Kainulainen:2009dw,Kainulainen:2010at,Kainulainen:2011zx}. The results were directly confronted to $N$-body simulations and shown to fit them very well up to a redshift of order 1.5, with the advantage of being given as a function of the relevant cosmological parameters. These fits can be employed to take into account the lensing extra scatter for any value of the cosmological parameters and also to model the lensing non-Gaussianity.

In this paper we propose instead to use the accurate determination of
the lensing moments of Ref.~\cite{Amendola:2013twa} to {\it measure} the cosmological parameters. As is often the case in cosmology, what was once a noise to be eliminated can become a signal when either data or modeling improve. Such idea in the present context has been first discussed in~\cite{Bernardeau:1996un,Hamana:1999rk,Valageas:1999ir} and later further developed in~\cite{Dodelson:2005zt}. We improve upon~\cite{Dodelson:2005zt} in two ways. First, we  use not just the variance of the lensing signal but the 3rd and 4th order moments as well. Second, we do not assume that the intrinsic SN variance is fixed, but we marginalize over it at every redshift bin independently. The first step boosts the sensitivity (this was first proposed in~\cite{Bernardeau:1996un}) while the second the robustness of the method and  allows us to show that a fundamental cosmological parameter, $\sigma_8$, can indeed be measured by LSST survey using SN lensing alone to within 3\%--7\%, a value that is competitive with usual methods based on cosmic shear, cosmic microwave background (CMB) or cluster abundance, and completely independent of these. In particular, it does not rely on measuring galaxy shapes (as cosmic shear) and is therefore immune to the systematics associated to the cross-correlation of intrinsic galaxy ellipticities. Also, it does not require to extrapolate the amplitude $\sigma_8$ from recombination epoch to today, as with the CMB technique, nor to make assumptions on the threshold of formation of structures that are needed when employing galaxy clusters. It is therefore a relatively direct measurement of $\sigma_8$ that can cross-check the results obtained via these other methods.

It is interesting to note that our proposal is essentially to carry out a one-point statistics on the supernova distribution on the Hubble diagram. This contrasts with other proposed methods which rely on two or higher point statistics such as that of~\cite{Cooray:2005yp}, where SN lensing and their inferred magnification was used as a tracer of dark matter clustering. In fact, by not relying on two-point statistics we avoid issues related to spatial correlations such as those arising from finite survey areas.

The main assumption that is needed for the SN lensing method is that the supernovae have an intrinsic magnitude distribution that is Gaussian, so that the entire non-Gaussianity can be attributed to lensing. In the future, this assumption can be directly tested by building a large calibrated sample of local supernovae. In principle, however, one could also include in the analysis an intrinsic non-Gaussianity and marginalize over it.

In order to make our method more directly applicable to future datasets, we will
base our estimation of $\sigma_8$ on the moments of the lensing distribution. One could use however also the full likelihood, again obtained via the sGL simulations. We will show however that a simplified likelihood based only on the first few moments is a very good approximation to the full likelihood.

The paper is organized as follows. In Section \ref{setup} we will describe the universe and lensing model adopted, and discuss the statistical properties of the lensing PDF as far as the central moments are concerned. In Section \ref{snalysis} we will examine the impact of lensing on the SN analysis, and in Section \ref{sec:constraining-s8} we will quantify how tightly can SNe constrain $\sig$. Finally, we will conclude in Section \ref{conclusions}. We will discuss some more technical details in the Appendices \ref{sec:app-full-cov} and \ref{sec:lecorre}, and provide in Appendix~\ref{sec:zits} redshift-dependent fits for the second-to-fourth central moments of the lensing PDF which are simplified versions of the original fits in~\cite{Amendola:2013twa}.

\section{Lensing moments}\label{setup}

In this Section we will first describe the model we will use to compute the moments of the lensing PDF. Then we will discuss the properties of the cumulants as far as the convolution of the lensing and supernova distributions is concerned.

\subsection{Universe and lensing model} \label{sec:model}

We will calculate the second-to-fourth central moments  $\mu_{2-4, \text{lens}}$ of the lensing PDF using the results of Ref.~\cite{Amendola:2013twa}.
There, accurate analytical fits as a function of $\{z,\,\sigma_{8},\,\Omega_{m0}\}$ were given for the broad ranges
$0 \leq  z \leq 3,\, 0.35 \leq \sigma_{8} \leq 1.25,\, 0.1 \leq \Omega_{m0} \leq 0.52$. The dependence of the lensing moments on other parameters (such as $\Omega_{k0},\, w_{0},\, n_{s}$) was shown to be almost negligible.
The results of Ref.~\cite{Amendola:2013twa} were obtained using the stochastic gravitational lensing (sGL) method introduced in Refs \cite{Kainulainen:2009dw,Kainulainen:2010at,Kainulainen:2011zx}.
The sGL method is based on (i) the weak lensing approximation and (ii) generating stochastic configurations of inhomogeneities along the line of sight.

Regarding (ii), the matter contrast $\delta_{M}(r,t)$ is modeled according to the so-called ``halo model'' (see, for example, \cite{Neyman:1952, Peebles:1974, Scherrer:1991kk, Seljak:2000gq, Ma:2000ik, Peacock:2000qk, Scoccimarro:2000gm, Cooray:2002dia}), where the inhomogeneous universe is approximated as a collection of different types of halos whose positions obey the linear power spectrum.
The halo model assumes that on small scales the statistics of matter correlations is dominated by the internal halo density profiles, while on large scales the halos are assumed to cluster according to linear theory.\footnote{In \cite{Amendola:2013twa} correlations in the halo positions were, however, neglected. As shown in~\cite{Kainulainen:2010at,Kainulainen:2011zx}, this should be indeed a good approximation for the redshift range of $z \lesssim 1$ in which we are mainly interested in this paper.}
The halos were modeled using the halo mass function given in Ref.~\cite{Jenkins:2000bv}, which has a good degree of universality~\cite{White:2002at}. The use of other mass functions such as the one given in Ref.~\cite{Courtin:2010gx} does not change substantially the results of our analysis.
The halo profiles are modeled according to the Navarro-Frenk-White (NFW) profile \cite{Navarro:1995iw}, which is able to model both galaxy-sized halos and superclusters with an appropriately chosen concentration parameter. The concentration parameter depends on the cosmology and we use the universal and accurate model proposed in Ref.~\cite{Zhao:2008wd}.

Regarding (i), the lens convergence $\kappa$ in the weak-lensing approximation is given by the following integral evaluated along the unperturbed light path~\cite{Bartelmann:1999yn}:
\begin{equation} \label{eq:kappa}
    \kappa(z_{s})=\int_{0}^{r_{s}}dr \, \rho_{MC} \, G(r,r_{s})\,\delta_{M}(r,t(r))
\end{equation}
where the quantity $\delta_{M}(r,t)$ is the local matter density contrast (which is modeled as described above), the density $\rho_{MC} \equiv a_0^3 \, \rho_{M0}$ is the constant matter density in a comoving volume, and the function $G(r,r_{s})=  \frac{4\pi G}{c^2 \, a}  \; \frac{f_{k}(r)f_{k}(r_{s}-r)}{f_{k}(r_{s})}$ gives the optical weight of a matter structure at the comoving radius $r$.
The functions $a(t)$ and $t(r)$ are the scale factor and geodesic time for the background FLRW model, and $r_{s}=r(z_{s})$ is the comoving position of the source at redshift $z_{s}$.
Also, $f_{k}(r)=\sin(r\sqrt{k})/\sqrt{k},\, r,\,\sinh(r\sqrt{-k})/\sqrt{-k}$ depending on the curvature $k>,=,<0$, respectively.
At the linear level, the shift in the distance modulus caused by lensing is expressed in terms of the convergence only:
\begin{equation} \label{eq:dm}
\Delta m(z) \simeq  5 \log_{10}\big [1-\kappa(z) \big] \simeq - \frac{5}{\log 10} \; \kappa(z) \,.
\end{equation}
%
Eq.~(\ref{eq:kappa}) connects the statistical distribution of matter to the
statistical distribution of convergences.
The sGL method for computing the lens convergence is based on generating random
configurations of halos along the line of sight and computing the associated
integral in Eq.~(\ref{eq:kappa}) by binning into a number of independent lens
planes.
A detailed explanation of the sGL method can be found
in~\cite{Kainulainen:2011zx,Kainulainen:2010at,Kainulainen:2009dw} and a
publicly available numerical implementation, the \tgl package, at
\href{http://www.turbogl.org/}{turbogl.org}.

Because of the theoretical approximations (weak lensing and halo model approximation) and modeling uncertainties (halo mass function and concentration parameter model) intrinsic in the sGL modeling, the results of Ref.~\cite{Amendola:2013twa} can be relied upon at the level of $\sim$10\%.

\subsection{Cumulants cumulate}

Observationally, the lensing PDF is convolved with the intrinsic standard-candle distribution. Now, there are fundamental statistical quantities
which are additive over convolutions: the cumulants. In other words,
if $X$ and $Y$ are two independent random variables, then the cumulants
$K_{i}(Z)$ of the convolution $Z\equiv X\star Y$ are just given
by $K_{i}(Z)=K_{i}(X)+K_{i}(Y)$.

Here and in the following, we will use $\mu_{i}',\mu_{i}$ and $K_{i}$
to denote respectively the $i$-th raw moment, central moment and cumulant. 
We will abuse this notation and often refer to the first raw moment (the mean) $\mu_{1}'$ using the notation for the first central moment $\mu_{1}$ which is identically zero.
If $p(x)$ is a PDF then
by defining the generating function
\begin{equation}
\phi(t)=\left \langle e^{tx} \right \rangle=\int e^{tx}p(x)\,\dd x \,,
\end{equation}
one has that the cumulants and moments are defined as
\begin{eqnarray}
\mu'_{n} & \,\equiv\, & \left. \frac{\dd^{n}\phi}{\dd t^{n}} \right |_{t=0} \,,\\
K_{n} & \,\equiv\, & \left. \frac{\dd^{n}\log\phi}{\dd t^{n}} \right |_{t=0} \,.
\end{eqnarray}

We will initially assume that all standard candles have a distribution
which is intrinsically Gaussian. By intrinsically we mean neglecting
any systematic effect (such as lensing itself) that might
distort this distribution. A Gaussian has only two non-zero cumulants,
namely $K_{1}=\mu'_{1}$ and $K_{2}=\mu'_{2}-\mu_{1}'^{2}=\sigma^{2}$.
The weak lensing PDF, in turn, has by definition an (almost) negligible mean, and thus $K_{1, \text{lens}}=\mu'_{1, \text{lens}} \simeq 0$ so that for the lensing
PDF the moments are all central moments.

Using the relation between cumulants and central moments, to wit
\begin{align}
K_2 &= \mu _2  \label{K2} \\
K_3 &=\mu _3 \\
K_4 &=\mu _4-3 \mu _2^2 
\end{align}
(where we used the fact that $\mu_{1}=0$) and the additivity of
the cumulants discussed above we can write the first central moments
of the convolved standard-candle PDF as (after straightforward manipulation)
\begin{align}
\mu_{2} & \;\equiv\;\sigma_{{\rm tot}}^{2}
\;=\;\sigma_{{\rm lens}}^{2}+\sigma_{\rm SN}^{2}\,,\label{eq:mu2}\\
\mu_{3} & \;=\;\mu_{3,{\rm lens}}\,,\label{eq:mu3}\\
\mu_{4} & \;=\;\mu_{4,{\rm lens}}+6\,\sigma_{{\rm lens}}^{2}\,\sigma_{\rm SN}^{2}
+  3 \, \sigma_{\rm SN}^{4} \,.\label{eq:mu4}
\end{align}
Here $\sigma_{\rm SN}^{2}$ is the variance of the unlensed standard candles, which is sourced by the intrinsic dispersion $\sint$ and by the observational error $\sigma_{\rm exp}$:
\begin{align} \label{errorsb}
\sigma_{\rm SN}^{2} = \sint^{2} + \sigma_{\rm exp}^{2} \,.
\end{align}
We will assume later on that $\sigma_{\rm exp}$ is negligible as compared to
$\sint$, such that $\sigma_{\rm SN} = \sint$. This assumption has no influence
on our method and results.

There are some immediate conclusions from the above relations. First, it means
that one can directly compare a measure of $\mu_{3}$, which is the unnormalized
skewness (not to be confused with the normalized skewness, defined by
$\mu_{3}/\sigma^{3}$), with a theoretical prediction such as the fitting
function for $\mu_{3,{\rm lens}}$ provided in Ref.~\cite{Amendola:2013twa}.
Second, it means that using other moments besides just $\mu_2$ one can break the
degeneracy between the cosmological parameters (basically $\Omega_{m0}$ and
$\sigma_8$) and the nuisance parameter $\sint$. Such a degeneracy was arguably
the most important limitation of a previous work in the
literature~\cite{Dodelson:2005zt}. We will come back to this possibility in
Section~\ref{sec:constraining-s8}.

Incidentally, since (for many SNe) the mean $\mu'_1$ gives a very precise
measurement of $\Omega_{m0}$ (roughly independent of lensing),  if one has
precise independent measurements of $\sigma_8$ one can use~\eqref{eq:mu2} to
make an estimate of the intrinsic dispersion $\sint(z)$ by subtracting from
$\sigma_{{\rm tot}}^{2}$ the variance due to lensing and the experimental error.
This can be an interesting result on its own, as it can help understand the
physics of SNe. We nevertheless do not explore this possibility further in this
work.


\section{Impact of lensing on SN analysis}
\label{snalysis}

The standard use of standard candles such as supernovae is to map the luminosity distance-redshift relation of the background FLRW model so as to constrain the content of the universe. From this point of view any fluctuation of the SN distance modulus, such as instrumental error and lensing, is seen as noise (completely opposite will be the approach followed in the next Section \ref{sec:constraining-s8}).
Therefore, as far as the standard SN analysis is concerned lensing has to be dealt with appropriately so as not to bias the parameter extraction of the {\em background} parameters.
It is important to account for both the skewness and the cosmology dependence of the lensing PDF~\cite{Amendola:2010ub}.
In particular, the redshift dependence of lensing can distort and rotate the confidence-level contours.

The inclusion of lensing complicates the usual analysis with the effect that a standard $\chi^{2}$ analysis is not longer possible or consistent.\footnote{Even without lensing the usual supernovae $\chi^{2}$ analysis, which is iterative and sets $\chi^{2} / {\rm d.o.f.} \equiv 1$ is not ideal as it can lead to biases and does not allow determination of $\sint$ or model-selection analysis. See \cite{Kim:2011hg,March:2011xa,Lago:2011pk} for more details.} The skewness of the lensing PDF violates indeed the Gaussian assumption, while the cosmology dependence of the normalization of the likelihood imposes that the latter is kept contrary to what is done in the usual $\chi^{2}$ analysis.
For example, if one is using a Gaussian likelihood with error depending on the theoretical parameters, then one cannot at the same time keep the parameter-dependent normalization factor and use a mock catalog with $\chi^{2}=0$. Indeed, a $\chi^{2}=0$ catalog is supposed to have the likelihood peaked at the chosen fiducial model, but this will not be the case as the parameter-dependent normalization factor will tilt the likelihood surface in a way which depends on the error.

\begin{figure*}[t!]
\begin{centering}
\includegraphics[width=2\columnwidth]{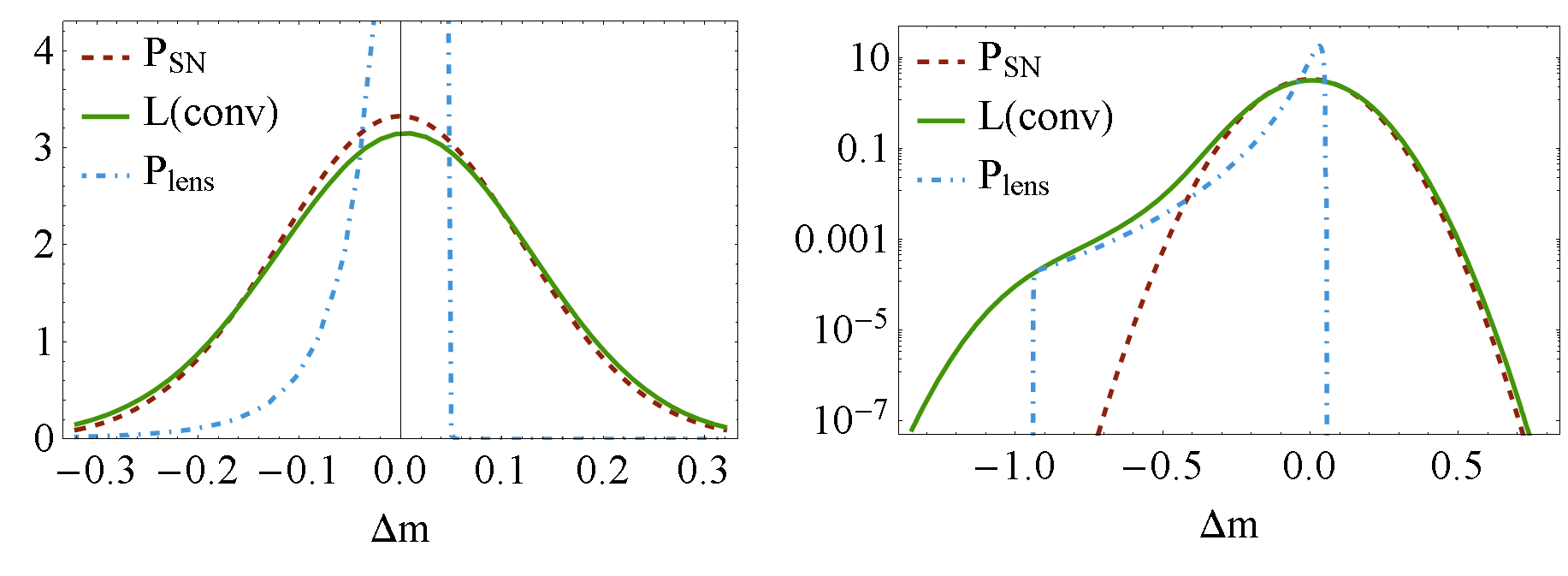}
    \caption{ Comparison of the original SN Gaussian PDF ($P_{\rm SN}$, dashed brown) with the lensed supernova PDF ($L$ of Eq.~(\ref{Li}), green), obtained by the convolution of the lensing PDF ($P_{\rm lens}$, dot-dashed blue) with $P_{\rm SN}$. We assume $\sigma_{\rm SN}=0.12$ mag, a redshift of $z=1$ and the WMAP9-only best-fit values as  fiducial model. As can be seen, the distortion of $L$ from Gaussianity is small around the peak, but it gets large in the high-magnification tail. For this plot we employ a cut $\kappa_{\rm cut} = 0.35$ (or $\Delta m = -0.94$ mag) for the lensing PDF, as discussed in Ref.~\cite{Amendola:2013twa}. See Section \ref{snalysis} for more details.
\label{fig:convolved-PDF}}
\end{centering}
\end{figure*}

A full likelihood analysis uses the lensing PDF, in our case either directly obtained from \href{http://www.turbogl.org/}{\tgl}or reconstructed using the log-normal approximation as proposed in Ref.~\cite{Amendola:2013twa}.\footnote{The lensing PDF may also be obtained using the ``universal'' lensing PDF of Ref.~\cite{Wang:2002qc}.} We denote the lensing PDF as $P_{\rm lens}(\Delta m,z, \omegam, \sig)$.
Within our approximations $P_{\rm lens}$ has negligible mean. The SN likelihood is then obtained by convolving the lensing PDF with the SN uncertainty distribution, which we assume to be Gaussian in the distance moduli and denote with $P_{\rm SN}(mean, sigma)$.
The latter uncertainty is sum of the intrinsic source brightness scatter and of the observational errors, as discussed in Eq.~(\ref{errorsb}).
The likelihood function for a single SN observation is then
\begin{align}
    L_{i}( \omegam, \sig, \xi)&=   \int \dd y \, P_{\rm lens}\Big(y,z_{i}, \omegam, \sig \Big) \; \times \label{Li}  \\
 \times \; & P_{\rm SN}\Big ( m_{t}(z_{i},\omegam)- m_{i} +\xi-y,\,\sigma_{{\rm SN},i} \Big)\,,\nonumber
\end{align}
where $z_{i}, m_{i}, \sigma_{{\rm SN},i}$ are the $i$th SN redshift, distance modulus and uncertainty, respectively. The quantity $m_{t}$ is the predicted distance modulus for a source at redshift $z_{i}$ in a flat $\Lambda$CDM model of matter density parameter $\omegam$:
\begin{align}
m_{t}(z, \omegam) = 5\log_{10}\frac{d_{L}(z)}{10\,\textrm{pc}}\,, \label{modulus}
\end{align}
where luminosity distance $d_{L}$ and Hubble rate $H$ are
\begin{align}
d_{L}(z) & =(1+z)\int_{0}^{z}\frac{\rd\bar{z}}{H(\bar{z})}\,, \\
\frac{H^{2}}{H^{2}_{0}} & =  \omegam (1+z)^{3}+\omegal  \,,
\end{align}
where $\omegal=1-\omegam$.

The quantity $\xi$ is an unknown offset sum of the supernova absolute magnitudes, of $k$-corrections and other possible systematics. Figure~\ref{fig:convolved-PDF} depicts the three distributions $P_{\rm SN}$, $P_{\rm lens}$ and $L_{i}$, where $P_{\rm lens}$ was modeled using the log-normal template proposed in Ref.~\cite{Amendola:2013twa}. As it can be seen, the distortion in $L_{i}$ from Gaussianity is small near the peak, but it gets large in the high-magnification tail. Finally, we define the total likelihood function as the product of all independent likelihood functions in the data sample, further marginalized over the unknown $\xi$:
\begin{equation}
    L_{\rm tot}(\omegam, \sig, \{\sintj\})=\int \dd \xi \; \Pi_{i}L_{i}( \omegam, \sig, \xi)\,,\label{fooltot}
\end{equation}
where we have explicitly stressed the dependence of $L_{\rm tot}$ on the intrinsic dispersion $\sint$, which we will leave as a set of free parameters $\{\sintj\}$, to be marginalized over in \emph{each redshift bin} $z_{j}$. Since $\xi$ is degenerate with $\log_{10}H_{0}$ we are effectively marginalizing also over the expansion rate of the universe.

As said earlier, SN observations are usually used to determine background parameters, in our case $\omegam$, but generally also $\Omega_{k0}, w_{0}, w_{a}$, etc. Therefore, the likelihood of Eq.~(\ref{fooltot}) has to be further marginalized over $\sig$, $L_{\rm bkg}=\int \dd \sig L_{\rm tot}$, and the question is how lensing affects $L_{\rm bkg}$. What we found is that while the effect of lensing can distort, increase and rotate the confidence-level contours of $L_{\rm bkg}$, it does not substantially bias the position of its maximum. The reason is twofold. First, the skewness induced by lensing is not sizable as far as the mean magnitudes are concerned and the variance added by lensing is subdominant with respect to the intrinsic dispersion of the supernovae, as shown by Figure~\ref{fig:convolved-PDF}. Second, the cosmology dependence of lensing will become less and less important as more precise measurements will restrict the available range of cosmological parameters within which the effect of lensing can vary.  Although sizable biases  were found in~\cite{Amendola:2010ub}, we now trace it to the toy model employed for the matter field (an universe populated with $~10^{14} M_\odot$ NFW halos). The much more realistic current modeling of \tgl produces weaker lensing effects.

The main focus of this paper is, however, not on $L_{\rm bkg}$ but on the signal hidden in the scatter of the SN distance moduli, in particular the information relative to $\sig$. Following this point of view, we will then use directly the likelihood $L_{\rm tot}$ of Eq.~(\ref{fooltot}).

\section{Constraining \boldmath $\sigma_{8}$ with SN data}
\label{sec:constraining-s8}

\subsection{The Method-of-the-Moments}

In this section we come back to the question of whether one can use
measurements of the observed distributions of standard candles at
different redshifts to constrain the statistics of matter inhomogeneities, in particular $\sig$. This
idea was first proposed by Ref.~\cite{Dodelson:2005zt} for supernovae,
but in that paper the authors focused mainly on using the additional
variance due to lensing. This is problematic, as in principle one
does not know what is the value of the intrinsic dispersion $\sigma_{{\rm int}}$ or even if
it is constant in redshift or not. Thus, even in the limit of perfect
modeling of instrumental errors and no extra systematics, a measurement
of a growing $\sint$ with $z$ could be attributed to
some sort of evolution effect of the supernovae rather than to lensing. The only way out
would be to reach a very good level of physical understanding of the
explosion process (and of other systematic effects) to be able to
accurately predict what $\sint(z)$ should look like.
In this work we break this degeneracy between the cosmological parameters and the nuisance parameter $\sint$ using other moments besides just $\mu_2$. In particular, we propose to measure the cosmological parameters $\{ \omegam, \sig \}$ and the distribution of $\sint(z)$ at the same time by using the information contained in the mean $\mu_{1}'$ and the first three central moments (which we will collectively refer to simply as $\mu_{1-4}$).

At this point we could just use the full likelihood $L_{\rm tot}$ of Eq.~(\ref{fooltot}), which automatically contains all the available lensing information.
However, we develop here an alternative approach which focuses, as outlined above, on the information carried by the lensing moments. There are mainly three reasons to do so:  (i) it is computationally faster as data can be binned in redshift\footnote{The full likelihood $L_{\rm tot}$ of (\ref{fooltot}) cannot be binned in redshift without losing information about moments above the second one.} and easily implemented in numerical codes; (ii) it does not require knowledge of the full lensing PDF and instead simply needs the theoretical prediction of the second-to-fourth lensing moments (available as analytical fits in~\cite{Amendola:2013twa}), which can directly be confronted with observations; and (iii) it will be essentially a $\chi^{2}$ approach (without need of convolutions), with all its advantages such as the ability to easily include nuisance parameters (for instance, describing some intrinsic non-Gaussianity of supernovae).
The disadvantage is that the likelihood gets somewhat more complicated as a result of the correlations between the moments, and in principle requires some information on high moments.

\begin{figure*}[t!]
\begin{centering}
    \includegraphics[width=2\columnwidth]{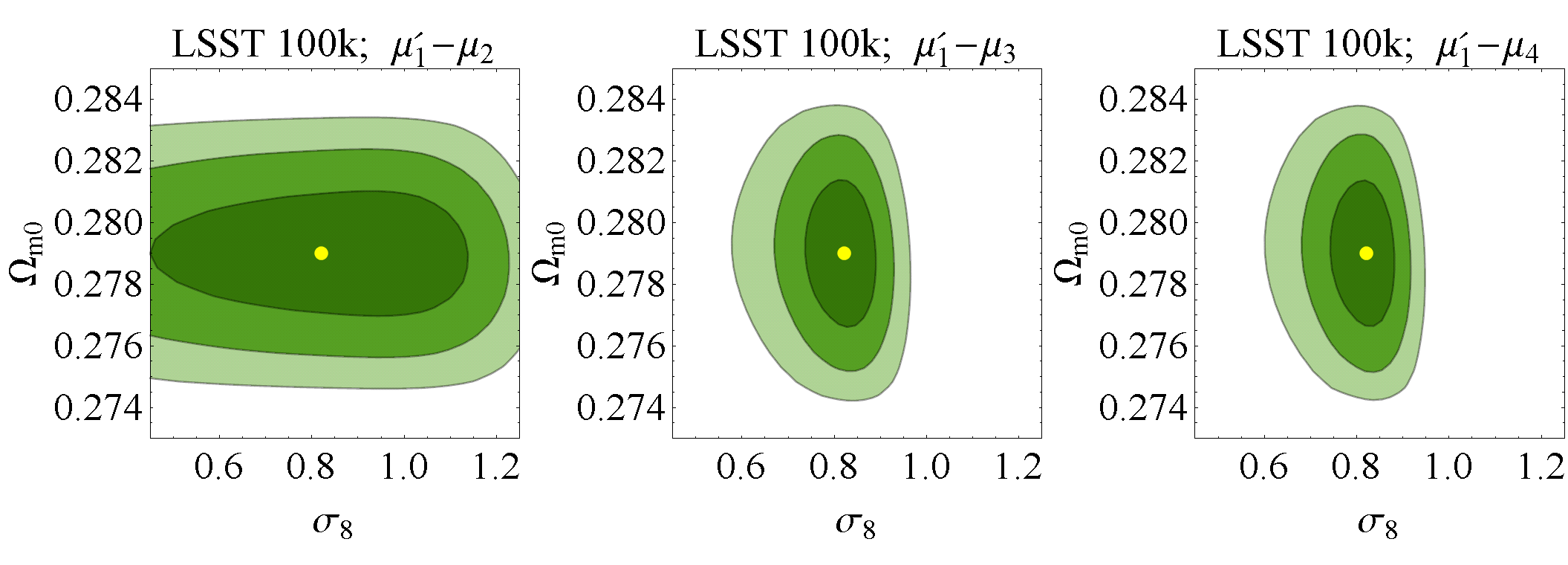}
    \caption{Comparison of the moment analysis using different combinations of  moments for the case of $10^5$ supernovae in LSST and assuming as usual a fiducial $\sigma_{\rm int} = 0.12$ mag. \emph{Left:} using $\mu'_1 - \mu_2$ only; \emph{middle:}  using $\mu'_1 - \mu_3$; \emph{right:} using $\mu'_1 - \mu_4$. We see that after marginalizing over $\sint$ in each redshift bin, very little information is gained from $\mu'_1 - \mu_2$ only, and one has to resort to at least the third moment. Note that although $\mu_4$ adds some extra information, most of it comes from $\mu'_1 - \mu_3$.
    \label{fig:mu1-2-3-4-comparison}}
\end{centering}
\end{figure*}

The idea is very simple: we build a likelihood at each redshift bin directly for
the first four moments $\mu_{1-4}$, to be called the \emph{method-of-the-moments}
(MeMo) likelihood:
\begin{align}
    &L_{\rm MeMo}(\omegam, \sig, \{\sintj\}) = \exp \left( - \frac{1}{2} \sum_{j}^{{\rm bins}} \chi_{j}^2 \right) \,,\label{Lmom} \\
    &\chi^2_j = \big(\boldsymbol{\mu}-\boldsymbol{\mu}_{\rm fid}\big)^t \;\Sigma_j^{-1}\; \big(\boldsymbol{\mu}-\boldsymbol{\mu}_{\rm fid}\big) \,, \label{chi2mom} \\
    &\boldsymbol{\mu} = \{ \mu_1',\,\mu_2,\,\mu_3,\,\mu_4 \} \,,
\end{align}
where the vector $\boldsymbol{\mu}(z_{j},\sigma_{8},\Omega_{m0},\{ \sintj \})$ is the theoretical prediction for the moments, and its second-to-fourth components are defined in~\eqref{eq:mu2}--\eqref{eq:mu4}. The mean $\mu_{1}'$ is the  theoretical distance modulus $m_{t}(z, \omegam)$ of~\eqref{modulus}.
The quantity $\boldsymbol{\mu}_{\rm fid}(z_{j})$ is the vector of fiducial or measured (sample) moments. In the former case it is $\boldsymbol{\mu}(z_{j},\sigma_{8},\Omega_{m0},\{ \sintj \})$ evaluated at the fiducial model, while in the latter case its components are:
\begin{align}
\mu_{1,{\rm fid}}'(z_{j}) & = \frac{\sum_{k} m_{k,j}}{N_j}  \,,  \label{sample1} \\
\mu_{i,{\rm fid}}(z_{j}) & = \frac{\sum_{k} [m_{k,j}- \mu_{1,{\rm fid}}'(z_{j})]^{i}}{N_j} \,, \label{samplei}
\end{align}
where $m_{k,j}$ are the SN distance moduli observed in the redshift bin centered at $z_{j}$. The covariance matrix $\Sigma$ is built using the fiducial (or observed) moments and therefore does not depend on cosmology (but it does on $z$).
This can be understood intuitively from (\ref{chi2mom}) as in this case  the $\chi^2_j$ function is minimized, as expected,  by $\boldsymbol{\mu}=\boldsymbol{\mu}_{\rm fid} $.
Consequently, the normalization factor (basically the determinant of $\Sigma$) is irrelevant and we have neglected it in (\ref{Lmom}).
The number of moments to be used in this analysis is in principle arbitrary as each new moment adds information. However, as will be shown below, for supernova almost all of the information is already included using $\mu_{1-4}$ (and a very good fraction of it already in $\mu_{1-3}$).

In the Gaussian limit of $L_{\rm tot}$, the covariance matrix relative to the redshift bin $z_{j}$ is simply
\begin{align}
\Sigma_{{\rm gau},j} = \frac{1}{N_j} \left (
\begin{array}{cccc}
 \sigma_j^2 & 0 & 0 & 0 \\
 0 & 2 \sigma_j^4 & 0 & 12 \sigma_j^6 \\
 0 & 0 & 6 \sigma_j^6 & 0 \\
 0 & 12 \sigma_j^6 & 0 & 96 \sigma_j^8 \\
\end{array}
\right) \label{sigmagau}
\end{align}
where $\sigma^{2}_j$ is the variance of the dataset, fiducial or observed, at the redshift bin centered on $z_j$. In the former case it is $\sigma^{2}_j = \sintj^{2}+ \sigma_{{\rm lens},j}^{2}$, with the latter evaluated at the fiducial flat $\Lambda$CDM model given by the 9-year WMAP-only best-fit values~\cite{Hinshaw:2012aka}. As discussed in Section~\ref{sec:model} we will consider only the dependence of lensing on $\{ z, \sigma_{8},\Omega_{m0} \}$. The fiducial values of the latter two are $0.821$ and $0.279$, respectively. The quantity $N_j$ is the number of SNe in the $j$-th redshift bin.
$\Sigma_{\rm gau}$ is a good approximation in the limit in which the deviation from the Gaussianity induced by lensing is small; in this case the covariance will be dominated by the Gaussian sampling variance. As we discuss in Appendix~\ref{sec:lecorre}, this is the case for standard candles with $z \lesssim 0.5$.

In the case of a general distribution the covariance matrix is more complicated. It can be written in a more compact way in terms of the cumulants:
\begin{widetext}
\begin{align}
& \Sigma_j =  \frac{1}{N_j} \times \label{sigmabadass} \\
&\vast(
\begin{array}{cccc}
 K_2 & K_3 & K_4 & 6 K_2 K_3+K_5 \\
 - & 2 K_2^2+K_4 & 6 K_2 K_3+K_5 & 12 K_2^3+14 K_4 K_2+6 K_3^2+K_6 \\
  - &  -& 6 K_2^3+9 K_4 K_2+9 K_3^2+K_6 & 72 K_3 K_2^2+18 K_5 K_2+30 K_3 K_4+K_7 \\
  - &  -&  -& 96 K_2^4+204 K_4 K_2^2+216 K_3^2 K_2+28 K_6 K_2+34 K_4^2+48 K_3 K_5+K_8 \\
\end{array}
\vast), \nonumber
\end{align}
\end{widetext}
where we denote by ``--'' the symmetric terms.
As can be seen, an accurate estimation of $\Sigma_j$ requires knowledge of all central moments up to $\mu_8$ (see Appendix \ref{sec:app-full-cov} for the relation between cumulants and moments). Nevertheless, as we will show, most of the information is contained in $\mu_{1-3}$, so in practice one would only need to go up to $\mu_6$. Moreover, as explained above, $\Sigma_j$ is to be evaluated at the fiducial values, so it is only necessary to know these moments at these fiducial values, and not as a function of $\{\sigma_{8},\,\Omega_{m0}\}$ (as carried out in~\cite{Amendola:2013twa}). As discussed before, the only nonzero cumulant of the Gaussian supernova PDF is $K_2$. Therefore $\Sigma$ can be evaluated, for each redshift bin, at the fiducial model simply by using $K_{2} = \sigma_{\rm lens}^2+\sint^{2}$ [see~\eqref{K2} and \eqref{eq:mu2}] and computing all the $K_{i}$ using \eqref{K2app}--\eqref{K8app} where the $\mu_{i} \rightarrow \mu_{i,{\rm lens}}$, i.e., the moments relative to the lensing PDF. Both Eqs (\ref{sigmagau}) and (\ref{sigmabadass}) were obtained using the \textsc{Mathematica} package \textsc{mathStatica} in the limit of large number of observations $N_{j}$ in each bin.

The MeMo likelihood of Eq.~(\ref{Lmom}) depends on the intrinsic dispersions $\{\sintj\}$, which are let free and independent in each redshift bin. In principle one has two choices in order to make forecasts of future constraints on $\{\sigma_{8},\,\Omega_{m0}\}$: either fixing $\sinti$ to a fiducial value (say $\sinti = 0.12$ mag, constant in redshift~\cite{Bernstein:2011zf,Abell:2009aa}) or leave it as a set of free parameters, to be marginalized over in each redshift bin. The former is computationally convenient as effectively reduces the dimension of the full likelihood by one. The latter is the more appropriate approach, and is the one recommended for analyzing real data. One has to choose the appropriate marginalization intervals for the flat priors to be used in (\ref{Lmom}). Although in principle these could be set as $\sinti \in \{0,\, \infty\}$, for all future SN catalogs here considered, the interval $\sinti \in \{0.1,\, 0.14\}$ mag proved sufficient.

\begin{figure}[t!]
\begin{centering}
    \includegraphics[width=.9\columnwidth]{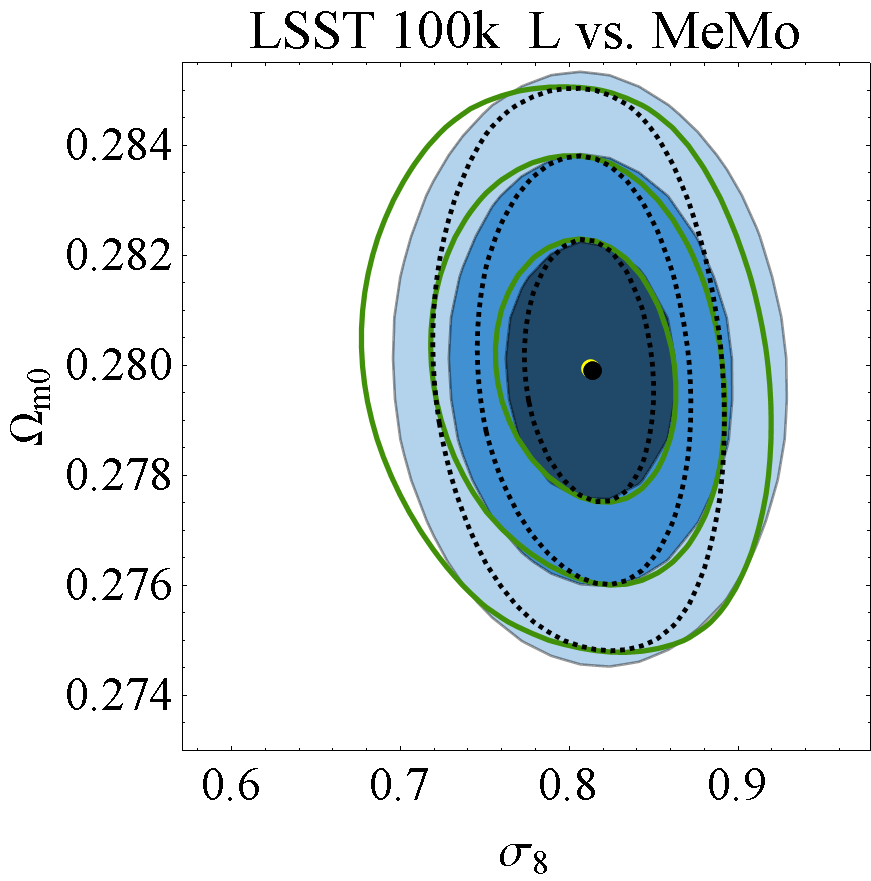}
    \caption{Comparison of the full likelihood analysis (colored contours) with the MeMo-likelihood one (solid green contours), using the full 4 dimensional covariance matrix of $\mu'_1 - \mu_4$ as given in Eq.~(\ref{sigmabadass}), for the case of $10^5$ supernovae in LSST and assuming a fixed $\sigma_{\rm int} = 0.12$ mag. Also shown is the MeMo results using only the diagonal part of the covariance matrix (black dotted contours). We see that the MeMo likelihood, using the mean and the second-to-fourth central moments, correctly reproduces the results relative to the full likelihood. Note that the fiducial of the MeMo contours was slightly changed in this plot to coincide with the one of the full likelihood and allow better comparison.
    \label{fig:mu1-4-cov-corr}}
\end{centering}
\end{figure}

\begin{figure}[t]
\begin{centering}
    \includegraphics[width=.9\columnwidth]{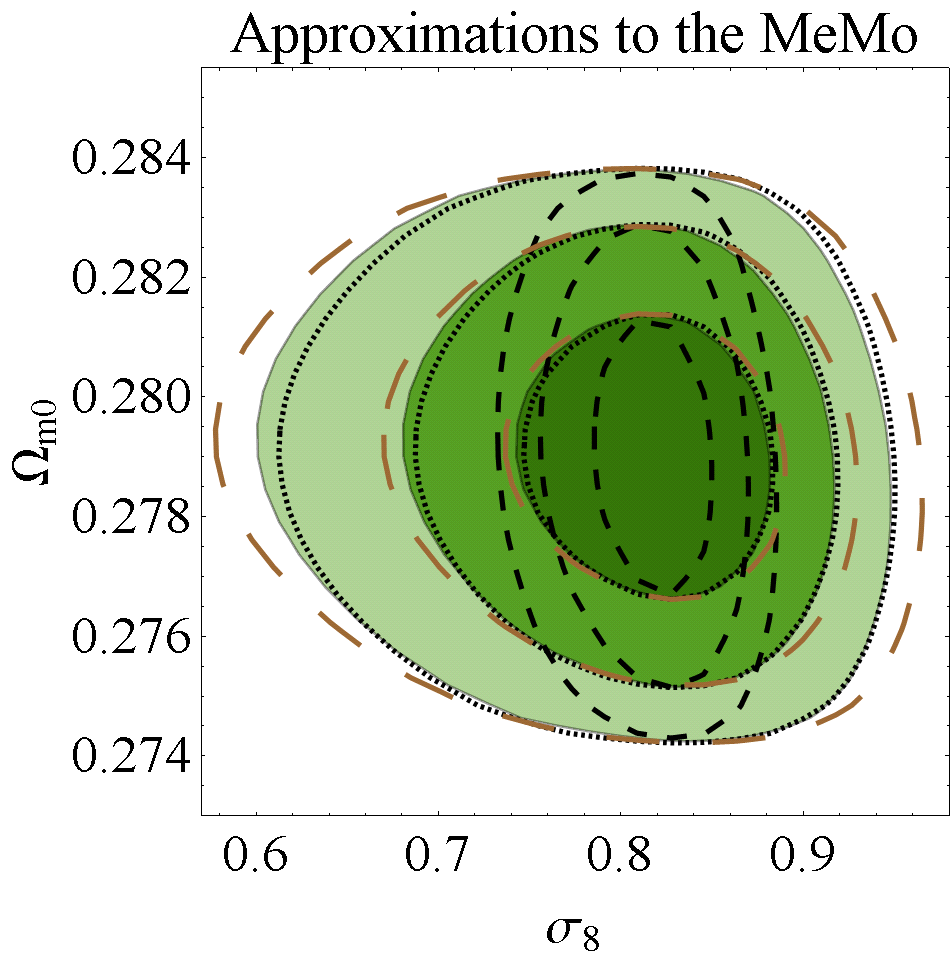}
    \caption{Comparison of the full covariance matrix~\eqref{sigmabadass} constraints (green contours) with 3 possible approximations: the Gaussian matrix~\eqref{sigmagau} (black  dashed contours), using only the diagonal part of~\eqref{sigmabadass} (black dotted contours) and using only $\mu'_1 - \mu_3$ (brown long-dashed contours) for the case of $10^5$ supernovae in LSST and marginalizing over all $\sintj$ (note that Figure~\ref{fig:mu1-4-cov-corr} in contrast assumes a fixed value of $\sigma_{\rm int}$). As can be seen, most of the information is already contained on the first 3 moments. Note also that the non-Gaussian diagonal matrix is a very good approximation after marginalization over $\sintj$, whereas the full Gaussian matrix underestimates $\delta \sigma_8$ to less than 50\% of the real value. \label{fig:cov-aproxs}}
\end{centering}
\end{figure}

Figure~\ref{fig:mu1-2-3-4-comparison} shows the comparison of the constraints using different combinations of moments as forecasted for the LSST 100k catalog, explained below in Section~\ref{sec:forecasts}. We see that after marginalizing over $\sint$ in each redshift bin, very little information is gained from $\mu'_1 - \mu_2$ only, and one has to resort to the third and (with marginal improvement) the fourth moment.
This was arguably the biggest caveat of the original work~\cite{Dodelson:2005zt}: in order to extract information from the variance only they had to assume the intrinsic dispersion to be well known. However, when one allows complete freedom for the different $\sintj$, almost no information can be collected from just the mean and the variance.

Figure~\ref{fig:mu1-4-cov-corr} confronts the full likelihood analysis with the MeMo, using $\mu'_1 - \mu_4$ assuming for computational convenience a fixed value of $\sint = 0.12$ mag in all bins. We see that by employing the full covariance matrix one can get a very good agreement between both methods. In what follows we will therefore stick to the MeMo method (using $\mu'_1 - \mu_4$) when deriving our forecasts.

The fact that, as seen from Figure~\ref{fig:mu1-2-3-4-comparison}, the fourth moment adds little constraining power to the analysis has the important consequence that the MeMo likelihood can be limited to mean, variance and skewness. This fact made more evident in Figure~\ref{fig:cov-aproxs} where we plot together both set of contours. This clearly makes the MeMo method more robust against the presence of SN outliers that could possibly bias the value of the higher moments. That being said, in what follows we will forecast results assuming all four moments are used.

\begin{figure*}[t]
\begin{centering}
\includegraphics[width=2\columnwidth]{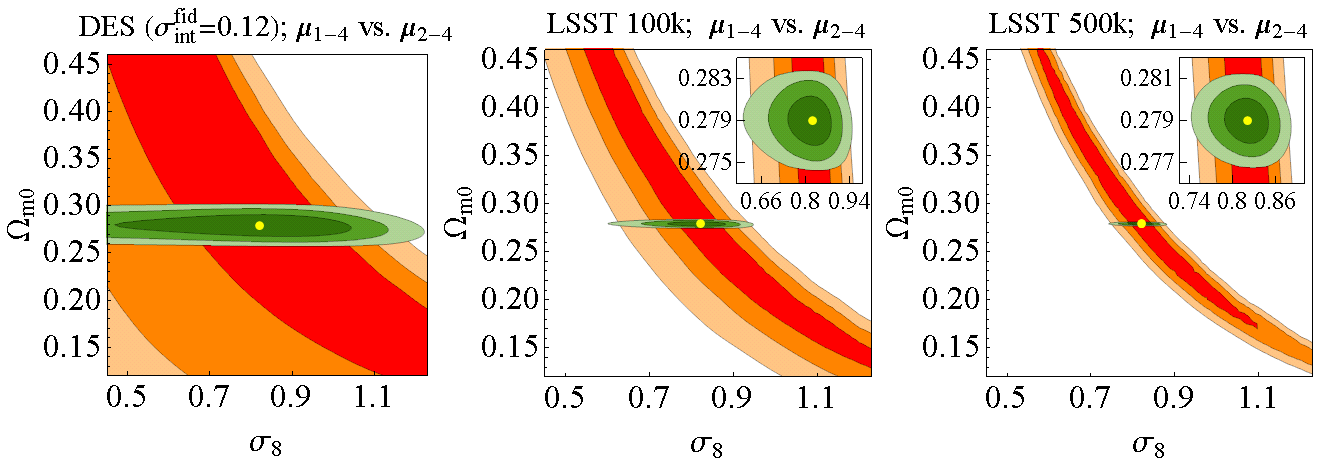}
    \caption{In green, forecast of cosmological constraints using the central moments
$\mu_{1}'$, $\mu_{2}$, $\mu_{3}$ and $\mu_{4}$ and supernovae from: \emph{(left)}
DES, if the scatter could be reduced to 0.12 mag (the predicted scatter given
in~\cite{Bernstein:2011zf} is somewhat larger);
    \emph{(middle)} LSST, with $10^5$ SNe, also assuming a fiducial intrinsic scatter of 0.12 mag (as predicted
    in~\cite{Abell:2009aa});
    and \emph{(right)} same for LSST but for $5 \times 10^5$ SNe. The red contours are the ``lensing-only'' likelihood,
see text.}
    \label{fig:full-constraints}
\end{centering}
\end{figure*}

Since, at least at first glance, the total convolved PDF is not largely dissimilar from a Gaussian (see Figure~\ref{fig:convolved-PDF}), one naturally wonders whether the full covariance matrix~\eqref{sigmabadass} can be approximated by the Gaussian one~\eqref{sigmagau}. We have carried out this test adding also a second approximation for \eqref{Lmom}, namely using only the (full, non-Gaussian) diagonal terms. Figure~\ref{fig:cov-aproxs} also compares the three $\mu'_1 - \mu_4$ covariances for the case of LSST 100k (see Section~\ref{sec:forecasts}). As can be seen, the non-Gaussian corrections to the errors are very relevant, and broaden the constraints on $\sig$ by a factor greater than two. However, these lensing corrections become negligible for smaller redshift as discussed in Appendix \ref{sec:lecorre}.
We also conclude that the diagonal part of $\Sigma$ carries almost the same information as the full matrix. Due to this finding, we provide in Appendix \ref{sec:lecorre} fitting functions with respect to lensing for the diagonal components of $\Sigma$.

It is worth pointing out at this point that the formalism developed in this
section could be generalized in a straightforward way so as to include a
fiducial $\mu_{i,{\rm int}}$ for $i=3,\,4$. In other words, it would be possible
to add some non-Gaussianity -- possibly due to the SN themselves and/or to other transients  -- also for the intrinsic supernovae PDF and
marginalize over these additional nuisance parameters.

\subsection{Constraints from future supernova surveys}\label{sec:forecasts}

In this Section we forecast the precision with which one can measure $\sigma_8$ with future supernova data. For this we study three different catalogs, based on 2 surveys: Dark Energy Survey (DES)~\cite{Bernstein:2011zf} and LSST~\cite{Abell:2009aa}. DES is already operational, and is expected to observe around 3000 SN during its observational cycle. LSST is expected to be operational by the end of the decade, and should observe a tantalizing amount of roughly 50000 SN per year. It is assumed in~\cite{Abell:2009aa} that these can be measured with an average scatter of $\sintj=0.12$ mag. We therefore assume in what follows a fiducial a scatter of $\sintj=0.12$ mag constant in all redshifts.\footnote{Note that only the fiducial is assumed constant. I.e., we still marginalize over $\sintj$ in each redshift bin, as described in Section~\ref{sec:constraining-s8}.}
The 3 cases we will consider here are (i) DES, with the distribution given in~\cite{Bernstein:2011zf} (note that the same work predicts a higher intrinsic scatter, varying in $z$ between 0.14 and 0.25 mag); (ii) LSST, with the redshift distribution given in~\cite{Abell:2009aa} and a total of $10^5$ supernovae (which we dub ``LSST 100k'' and which should correspond to 2 years of observation); and (iii) same as (ii) but for a total of $5\times10^5$ SNe (``LSST 500k'', corresponding to the full 10-year survey).

Figure~\ref{fig:full-constraints} depicts the constraints on $\{\omegam, \sig \}$ that can be obtained with the future supernova surveys described above. From left to right, we use DES, LSST 100k and LSST 500k. The green contours are the final $1,\,2,\,3\sigma$ constraints, using~\eqref{Lmom} and \eqref{sigmabadass}. Note that LSST  is thus forecasted to measure $\sigma_{8}$ with error bars of around 0.057 with $10^5$ SNe (0.023 with $5\times 10^5$), which is around the interesting $7\%$ ($3\%$) level. For DES, however, the statistical information is underwhelming, and one will not gain good knowledge on $\sigma_{8}$, even though we assumed an intrinsic scatter smaller than the one estimated by the DES collaboration~\cite{Bernstein:2011zf}. The red contours, mostly of bookkeeping interest, depict the constraints provided by the lensing effect only (without using information about the mean). In other words, from the non-Gaussianity introduced by lensing on the final PDF. More precisely, they are the constraints obtained substituting the first row and column of~\eqref{sigmabadass} by zeros. Note that the red and green contours are almost perpendicular, which shows that the final constraints can be approximated by taking independently mean (used in standard SNe analyses) and lensing effects.

\begin{figure}[t]
\begin{centering}
    \includegraphics[width=0.95\columnwidth]{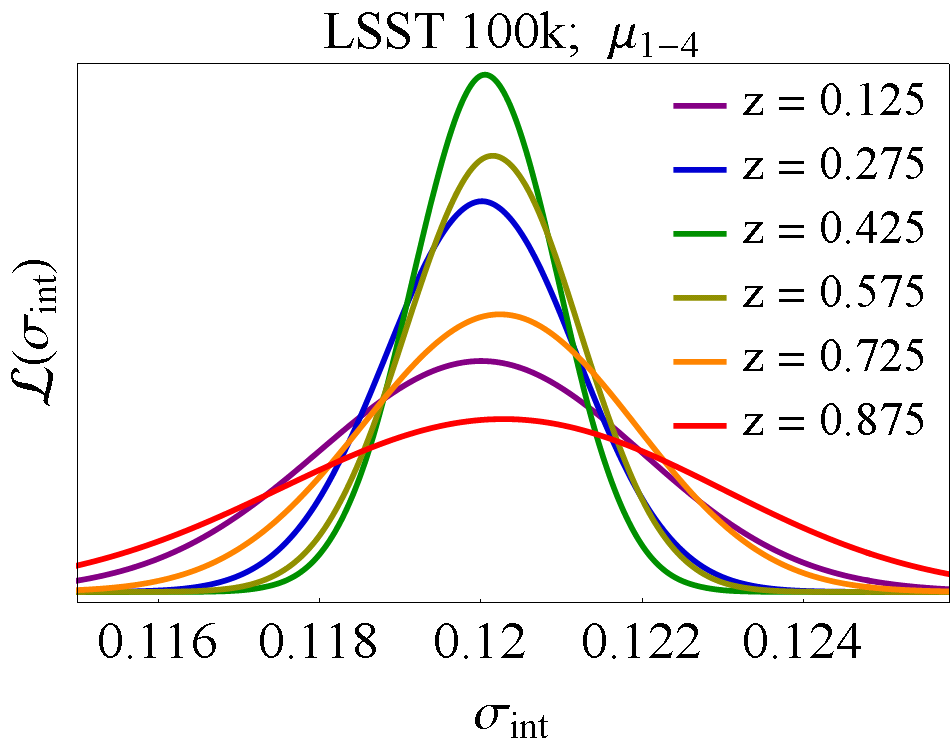}
    \caption{The posterior distribution of $\sint$ (in magnitudes) in different redshift bins of width $\Delta z = 0.05$ using LSST 100k. Although this catalog has 18 such bins, we depict only 6 for clarity. For low-$z$ lensing is small and the posterior is wide. For intermediate $z$ the constraints are the tightest (they peak at around $z=0.5$), because that is where there are more SNe. For the highest $z$-bins the number of forecasted supernovae decreases and the peaks broaden again.}
    \label{fig:sint-posterior}
\end{centering} %
\end{figure}

\begin{figure*}[t]
\begin{centering}
    \includegraphics[width=\columnwidth]{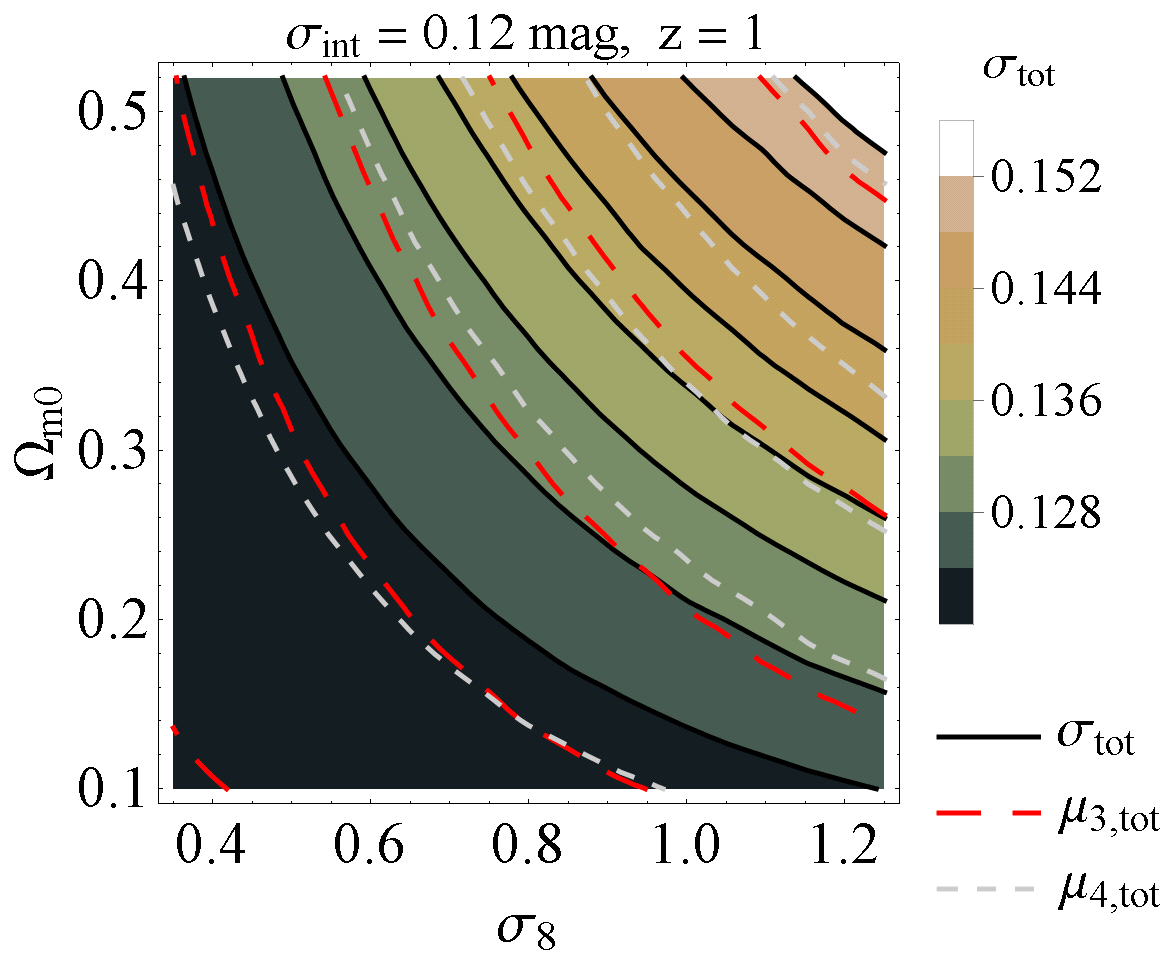}
    \;\;\includegraphics[width=\columnwidth]{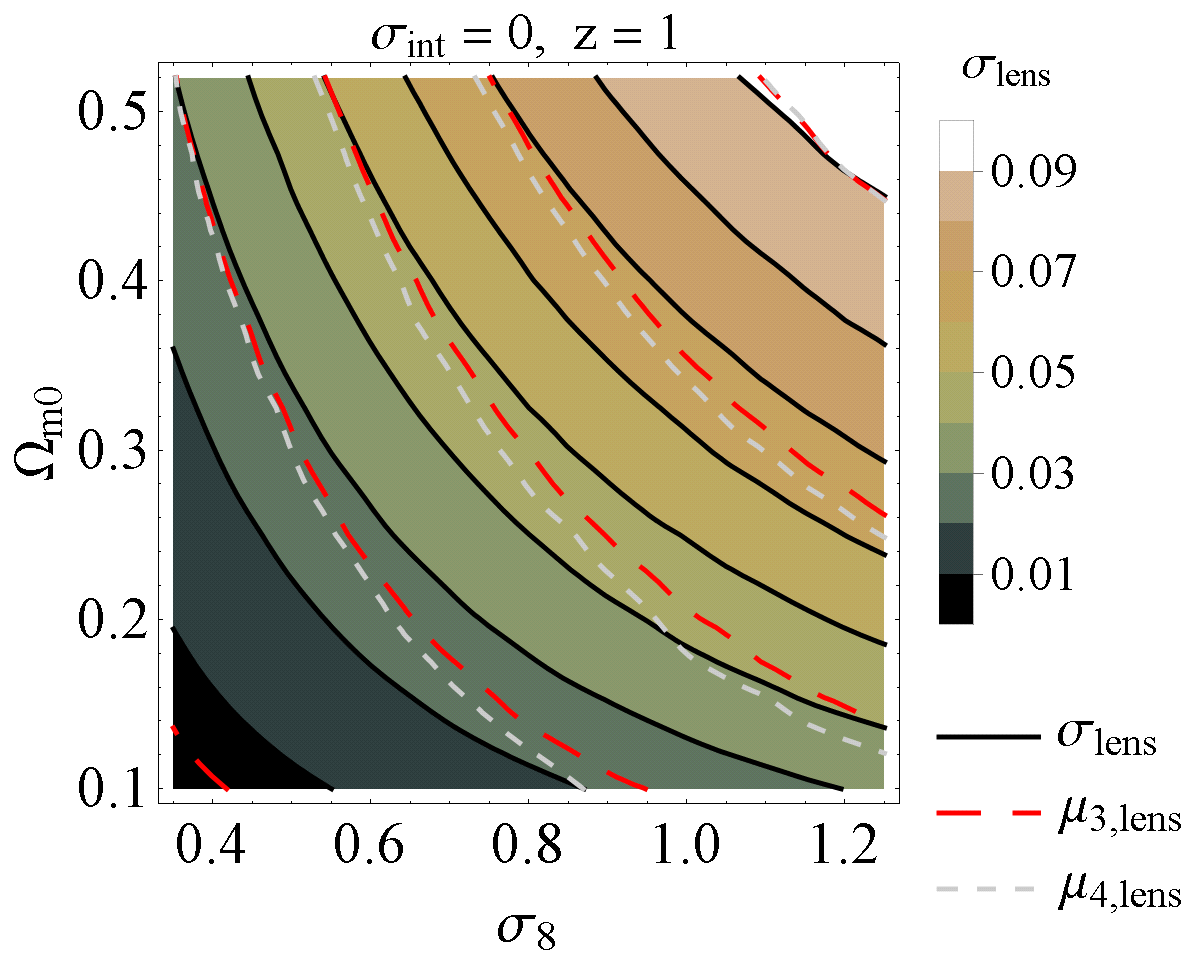}
    \caption{Constant central-moment contours, both for the total convoluted PDF (left) and for the lensing-only PDF (right). For each moment all contour levels are separated by a constant amount. Note that there is a near but not exact degeneracy, which explains why the higher-moment-only contours (i.e.~without the mean) have a finite area (as in the case of LSST in the range of parameters considered -- see Figure~\ref{fig:full-constraints}).}
    \label{fig:mu-contours}
\end{centering} %
\end{figure*}

Besides constraints in $\sigma_{8}$, the central moments (specially $\mu_2$) also give information on the intrinsic scatter of SNe in each redshift bin. Figure~\ref{fig:sint-posterior} depicts the achievable precision in measuring $\sintj$ in different redshift bins of width $\Delta z = 0.05$ using LSST 100k. Although this catalog has 18 such bins, we only depict 6 of them for clarity. The posteriors are all normalized to have unit area. The precision with which one can measure the different $\sintj$ varies roughly from 0.01 to 0.03 mag, depending on the redshift.

To illustrate the physics behind the red (``lensing-only'') contours of Figure~\ref{fig:full-constraints}, it is useful to depict the contours of constant value for the different central moments. This is done in Figure~\ref{fig:mu-contours}, both for the total convoluted PDF (left) and for the lensing-only PDF (right). Note that there is a near but not exact degeneracy, which explains the fact that although very broad, the lensing-only contours are in fact closed.

\section{Conclusions}
\label{conclusions}

In this paper we developed a method to measure $\sigma_8$ by exploiting the
lensing effects of matter clustering along the line of sight of SNe, extending
the results of~\cite{Dodelson:2005zt}. We have shown that
one can obtain interesting constraints in a survey with one or few hundred thousands supernovae,
as in the LSST survey.
In particular, we find that $\sigma_8$  can be estimated to within 3\% if 500,000 supernovae
with average magnitude error 0.12 are collected.
This method is independent of and complementary to the standard methods using
CMB, cosmic shear or cluster abundance and bypasses the need to assume a constant intrinsic variance
as in~\cite{Dodelson:2005zt}. We still assume that the SN magnitude distribution
does not have an intrinsic non-Gaussianity, although in principle one can marginalize over this extra
parameter.

Instead of employing the full likelihood, we have shown that the method of moments (MeMo) provides
a very good approximation and is much faster to implement. Even the simplest case of diagonal approximation works with sufficient accuracy (we provide fitting functions in Appendix \ref{sec:lecorre}).
We have shown that it is enough to use the first three moments (loosely speaking, mean, variance and skewness) so as to capture the non-Gaussian information available in the full likelihood.
This should make the MeMo method robust against the presence of SN outliers that could possibly bias the value of higher moments.
The needed moments as a function of cosmological parameters have been already derived in \cite{Amendola:2013twa}.

One could at first be suspicious about the feasibility of measuring the full covariance matrix~\eqref{sigmabadass}, as it assumes we are able to measure all moments up to the $8^{\rm th}$. Nevertheless, we have two reasons to believe this is not as hard as it seems. First, similarly to what was discussed in part I~\cite{Amendola:2013twa},
using the results of~\cite{Takahashi:2011qd} we evaluated the dependence of the first 8 moments on the point at which we cut the tail of the $\kappa$PDF. The dependence is not very strong and saturates somewhere between $\kappa_{\rm cut}$ of 0.2 and 0.7, depending on the moments. This range reflects the fact that higher moments indeed require a higher value of $\kappa_{\rm cut}$ for more precise results. However: (i) even for the $8^{\rm th}$ moment using $\kappa_{\rm cut} = 0.35$ would result in only a $20\%$ error in the value of $\mu_8$; (ii) as we have shown in the text, a very good estimate of $\sigma_8$ can be achieved neglecting the $4^{\rm th}$ moment entirely, and thus dropping the need for either the $7^{\rm th}$ and $8^{\rm th}$ moment for the computation of the covariance matrix. For $z>1$ this issue could become important, not so much because one needs a slightly larger $\kappa_{\rm cut}$ but mainly because turboGL in its current form starts to deviate from the numerical simulations due to medium and strong-lensing corrections.

Finally, we note that this method can be extended to other parameters, for instance to the growth function of matter perturbations. This will be explored in future work.

\begin{figure}[t]
\begin{centering}
    \includegraphics[width=.95\columnwidth]{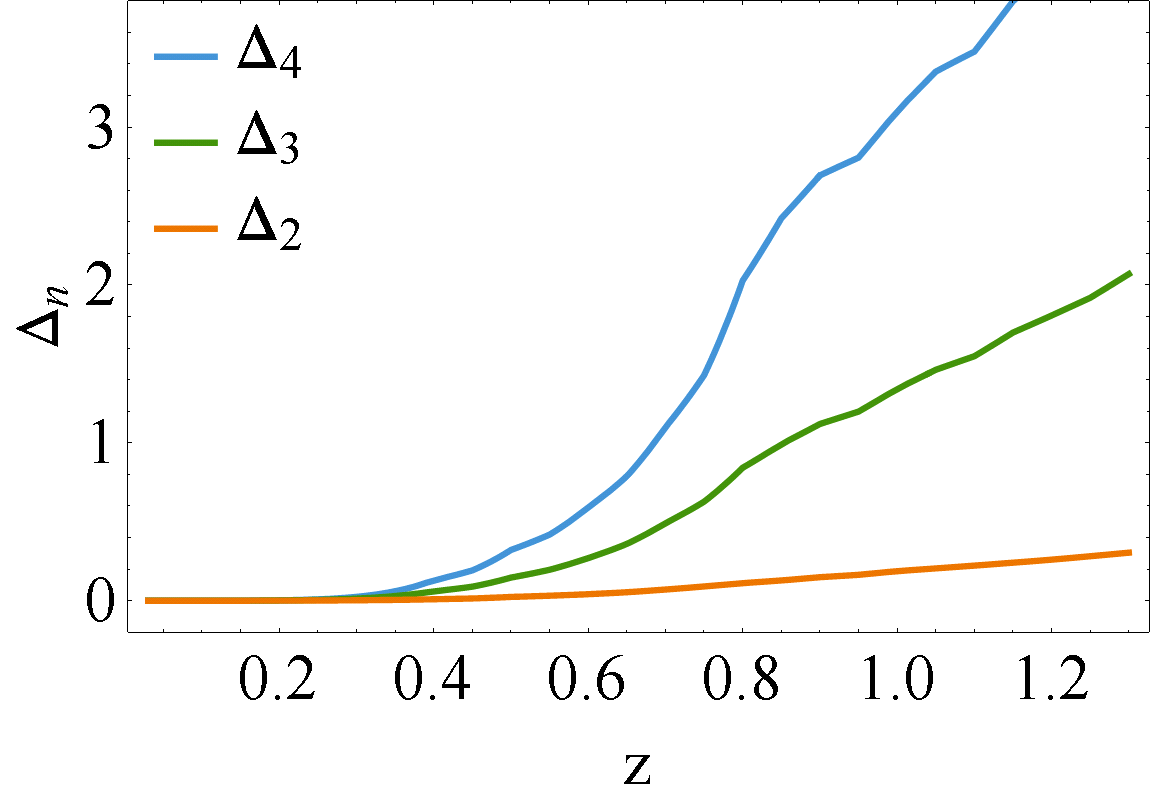}
    \caption{Value of the correction factors $\Delta_{n}$ for the variance of the moments using the WMAP9 fiducial values of $\{\sig \simeq0.821, \,\omegam \simeq 0.279\}$. \emph{Bottom to top:} $\Delta_{2}$, $\Delta_{3}$ and $\Delta_{4}$. For supernova forecasts in the usual domain $0<z<1.2$ these corrections can be large, specially for $\sigma^2(\mu_4)$. For  higher redshifts and for higher values of $\{\sig, \,\omegam\}$ they become even larger. \label{fig:Delta-n}}
\end{centering}
\end{figure}

\begin{acknowledgments}
It is a pleasure to thank Stephan Hilbert, Martin Makler, Bruno Moraes, Ribamar
Reis, Peter Schneider, Brian Schmidt and Ryuichi Takahashi for fruitful discussions. LA and VM acknowledge support from DFG through the TRR33 program ``The Dark
Universe''. MQ is grateful to Brazilian research agencies CNPq and
FAPERJ for support and to ITP, Universität Heidelberg for hospitality
during part of the development of this project.
\end{acknowledgments}

\appendix

\section{Higher lensing moments}\label{sec:app-full-cov}

As discussed in Section~\ref{sec:constraining-s8}, to compute the covariance matrix for the MeMo method one needs an estimation of all the central moments up to $\mu_8$ at the given fiducial cosmology. We thus extend here the relations between cumulants and central moments all the way to $K_8$:
\begin{align}
K_2 &= \mu _2  \label{K2app} \,, \\
K_3 &=\mu _3 \,, \\
K_4 &=\mu _4-3 \mu _2^2 \,, \\
K_5 &=\mu _5-10 \mu _2 \mu _3 \,, \\
K_6 &= 30 \mu _2^3-15 \mu _4 \mu _2-10 \mu _3^2+\mu _6 \,, \\
K_7 &=210 \mu _3 \mu _2^2-21 \mu _5 \mu _2-35 \mu _3 \mu _4+\mu _7 \,, \\
K_8 &=-630 \mu_2^4+420 \mu _4 \mu _2^2+560 \mu _3^2 \mu _2 \nonumber \\
& -28 \mu _6 \mu _2 -35 \mu _4^2-56 \mu _3 \mu _5+\mu _8 \,.  \label{K8app}
\end{align}
In part I of this work~\cite{Amendola:2013twa} we provided fits for $\mu_{2,{\rm lens}}$, $\mu_{3,{\rm lens}}$ and $\mu_{4,{\rm lens}}$. For the present work we have also computed the values of $\mu_{5,{\rm lens}}$--$\mu_{8,{\rm lens}}$. Surprisingly, it turns out even these very high moments, when computed with \tgl are still in very good agreement with $N$-body simulations such as~\cite{Takahashi:2011qd}.

\section{Lensing corrections for the covariance matrix}\label{sec:lecorre}

Although we have shown in Figure~\ref{fig:cov-aproxs} that the Gaussian covariance matrix~\eqref{sigmagau} is not a very good approximation for the supernova catalogs here considered, it can actually be used up to intermediate redshifts, for which lensing and thus also the non-Gaussianity are small. We compute the corrected values of the variance for a given $\mu_n$ in terms of normalized correction factors $\Delta_n$, defined by
\begin{align}
    \sigma_{\mu_{n}}^{2} \equiv \sigma_{\mu_n,\,{\rm gau}}^{\,2} (1+\Delta_n),
    \label{eq:var-mun-corr}
\end{align}
so that $\Delta_n = 1$ represents a corrected variance which is twice the Gaussian ones in~\eqref{sigmagau}. Figure~\ref{fig:Delta-n} shows this for the diagonal terms of~\eqref{sigmabadass}. Note that the corrections start to be relevant for $z \gtrsim 0.5$, especially for the higher moments.


\begin{figure}[t!]
\begin{centering}
    \includegraphics[width=.97\columnwidth]{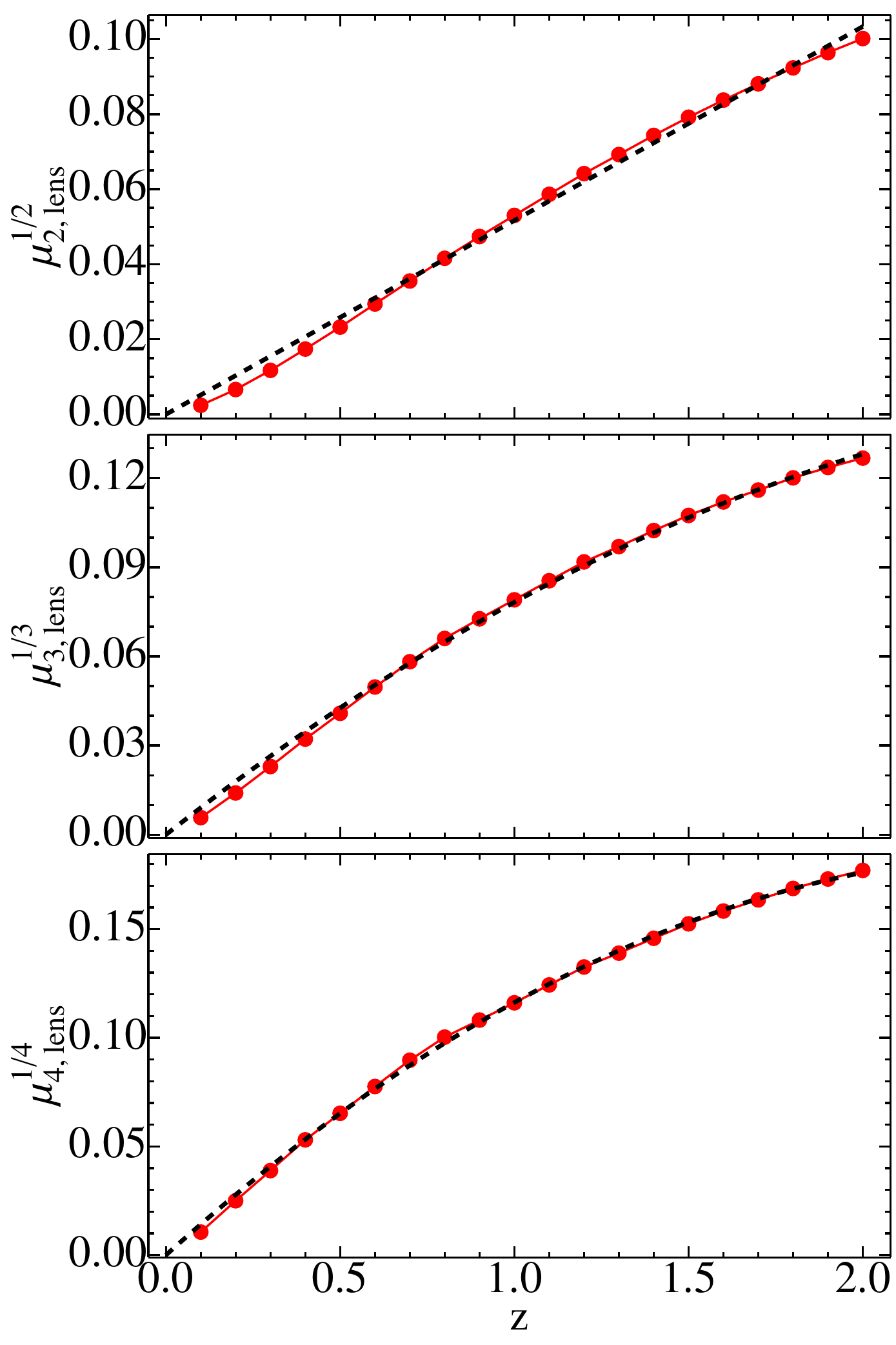}
    \caption{Comparison between the simple 
    fits \eqref{zf2}--\eqref{zf4} of the lensing moments (black dashed line) and the \tgl output (red connected dots). The fiducial cosmology used is the best fit obtained by the Planck Collaboration. All values are in magnitudes.}
    \label{fig:zfits}
\end{centering}
\end{figure}

As shown in Fig.~\ref{fig:cov-aproxs}, the diagonal of the covariance matrix $\Sigma$ of Eq.~(\ref{sigmabadass}) is sufficient to obtain an accurate forecast of the constraints on $\{ \omegam, \sig \}$. Consequently, we provide fitting functions with respect to redshift in the range $0\le z\le 1.5$ (average error below 5\%) for the diagonal components of $\Sigma$ (note that this assumes $\sint = 0.12$ mag):
\begin{align}
10^{4} \Sigma_{11} & =  144 + 43.8 z^3 - 30.1 z^4 + 6.64 z^5  , \\
10^{5} \Sigma_{22} & = 41.5 -0.88 z +35 z^3 -11.7 z^4 , \\
10^{5} \Sigma_{33} & =  1.79+7.77 z^4 -3.79 z^6 +0.44 z^9 ,  \\
10^{5} \Sigma_{44} & = 0.414 -0.34 z^4 +12.8 z^5 -14.8 z^6 +4.71 z^7,
\end{align}
which can be directly used to calculate the constraints on $\{ \omegam, \sig \}$ (but possibly also on other parameters).

$\,$\vspace{0.0cm}$\,$

\section{Simplified Redshift-dependent fits}\label{sec:zits}

In this Section we will provide simple redshift-dependent fits for the second-to-fourth central moments of the lensing PDF. They have been computed using the mass function of Ref.~\cite{Courtin:2010gx} and assuming as fiducial cosmology the best fit obtained by the Planck Collaboration when assuming a spatially flat  $\Lambda$CDM model and fitting to observations of the cosmic microwave background and baryon acoustic oscillations~\citep[][(Table 5, last column)]{Ade:2013zuv}.
These fits are complementary to the flexible cosmology-dependent fits presented in part I of our present investigation~\cite{Amendola:2013twa}, and are meant to be used when the dependence on $\{\sigma_{8},\,\Omega_{m0}\}$ is not important and when the smaller redshift range $0\le z\le 2$ is enough:
\begin{align}
    \sigma_{\rm lens}(z) & =  0.052 z  \, ,  \label{zf2} \\
    \mu_{3,{\rm lens}}^{1/3}(z) & = 0.092 z - 0.014 z^{2}  \, , \\
    \mu_{4,{\rm lens}}^{1/4}(z) & =  0.14 z - 0.028 z^{2} \, .  \label{zf4}
\end{align}
%
The comparison of these fits with the full numerical calculations is shown in Fig.~\ref{fig:zfits}.


The first of the above fits can also be more directly compared to other independent estimates of the additional lensing variance. For instance, early Supernova Legacy Survey data was used to estimate (again, in magnitudes) $\,\sigma_{\rm lens}(z)  =  (0.055 \pm 0.04) z$~\cite{Jonsson:2010wx} and $\,\sigma_{\rm lens}(z)  =  (0.054 \pm 0.024)z$~\cite{Kronborg:2010uj}, while recent independent theoretical computations estimated $\sigma_{\rm lens}(z)  =  0.056 z$~\cite{BenDayan:2013gc}.

\bibliography{cosmo-lensing}

\end{document}